\documentclass[aps,prl,twocolumn,showpacs,groupedaddress]{revtex4}  
\usepackage{graphicx}  
\usepackage{dcolumn}   
\usepackage{amssymb,amsmath}   

\begin{document}

\title{Bose condensation far from equilibrium}
\author{J{\"u}rgen Berges}
\author{D\'enes Sexty}

\affiliation{Universit\"at Heidelberg, Institut f{\"u}r Theoretische Physik, Philosophenweg 16, 
69120~Heidelberg, Germany\\
and ExtreMe Matter Institute (EMMI),
GSI Helmholtzzentrum f\"ur Schwerionenforschung GmbH, 
Planckstra\ss e~1, 64291~Darmstadt, Germany
}

\begin{abstract}
The formation of Bose condensates far from equilibrium can play an important role in our understanding of collision experiments of heavy nuclei or for the evolution of the early universe. In the relativistic quantum world particle number changing processes can counteract Bose condensation, and there is a considerable debate about the relevance of this phenomenon in this context. We show that the involved question of Bose condensation from initial over-population can be answered for the example of scalar field theories. Condensate formation occurs as a consequence of an inverse particle cascade with a universal power-law spectrum. This particle transport towards low momenta is part of a dual cascade, in which energy is also transfered by weak wave turbulence towards higher momenta. To highlight the importance of number changing processes for the subsequent decay of the condensate, we also compare to non-relativistic theories with exact number conservation. We discuss the relevance of these results for nonabelian gauge theories. 
\end{abstract}

\maketitle

Seventy years after the prediction of Bose and Einstein for a new state of matter of a dilute gas of weakly coupled bosons the first gaseous condensate was produced using ultracold rubidium atoms~\cite{NP}. By now a very detailed understanding exists, both experimentally as well as theoretically, about how a condensate emerges by  decreasing the temperature going through a sequence of close-to-equilibrium states. Much less is known about Bose condensation far from equilibrium. Most importantly, this concerns open questions about condensation dynamics in relativistic quantum field theories including particle creation and annihilation processes. Particle number changing processes might change the situation dramatically as compared to the number conserving non-relativistic counterparts. This can play a crucial role in our understanding of far-from-equilibrium stages of the early universe~\cite{PH} as well as for related questions in relativistic collision experiments of heavy nuclei~\cite{HIC}.

Recently, it has been argued that Bose condensation in relativistic quantum field theories occurs for a generic class of far-from-equilibrium situations. For the example of a plasma of quarks and gluons in the early stages of a heavy-ion collision at sufficiently high energies, the gluon density may be parametrically large compared to the thermal equilibrium value~\cite{Blaizot:2011xf}. As a consequence of the initial over-population, where the gluon occupation number at a large characteristic momentum scale $Q_s$ is parametrically of the order of the inverse gauge coupling $1/\alpha_s(Q_s)\gg 1$, condensation is suggested to occur. Similar situations arise also in the context of scalar inflaton dynamics in the early universe following the violent process of preheating~\cite{PH}. Here a nonequilibrium instability, such as parametric resonance in chaotic inflationary scenarios or a tachyonic instability, leads to over-population of typical modes with occupancies inversely proportional to a small quartic self-coupling $\lambda \ll 1$. Despite the weak coupling, the parametrically large occupancies lead to strong correlations such that the problem is non-perturbative. 

In this work we show that Bose condensation far-from-equilibrium generically occurs from initial over-population as a consequence of a particle cascade towards low momenta. We consider relativistic scalar field theories where the occupation number at a large characteristic momentum scale, $n(Q_s)$, starts out proportional to the inverse quartic self-coupling, $1/\lambda \gg 1$. Remarkably, as a consequence of this non-perturbative occupancy the condensation dynamics becomes universal. For a large class of models, independently of the value of the (small) coupling or mass or (sufficiently large) cutoff scale, the same phenomenon is observed. A dual cascade develops: for momenta $p \gtrsim Q_s$ a power-law distribution 
\begin{equation}
n(p\gtrsim Q_s) \, \sim \, \left(\frac{Q_s}{p}\right)^{d-3/2}
\label{eq:pertscaling}
\end{equation}
signals the direct energy cascade towards higher momenta known from the theory of weak wave turbulence in $d$ spatial dimensions~\cite{Kolmogorov,Micha:2004bv}. This energy cascade goes along with an inverse particle cascade towards the infrared with the low-momentum distribution 
\begin{equation}
n(p\lesssim Q_s) \, \sim \, \left(\frac{Q_s}{p}\right)^{d+1}
\label{eq:nonpertscaling}
\end{equation}
known from strong turbulence applications to early-universe inflaton dynamics~\cite{Berges:2008wm,Berges:2008sr}. Corresponding power-law regimes with directed fluxes have also been found in non-relativistic systems of cold atoms~\cite{Scheppach:2009wu,BS}.

We find that the transport of particles from the infrared cascade leads to condensation of the zero momentum mode of the anti-commutator expectation value of two Heisenberg field operators $\hat{\phi}(t,\vec{x})$, 
\begin{equation}
F(t,t';\vec{x} - \vec{y}) \, = \, \left\langle \left\{ \hat{\phi}(t,\vec{x}), \hat{\phi}(t',\vec{y}) \right\} \right\rangle \, .
\end{equation} 
In spatial Fourier space, $F(t,t';p) = \int \mathrm{d}^d p/(2\pi)^d \exp(i\vec{p}\cdot\vec{x})\, F(t,t';\vec{x})$ with $p\equiv |\vec{p}|$ can be decomposed into a (time-dependent) condensate fraction $\sim \phi_0^2(t)$ and a non-condensate fraction involving the distribution function $n_p(t) \equiv n(t,p)$ as well as the dispersion $\omega_p(t)$,
\begin{equation}
F(t,t; p) \, = \, \frac{1}{\omega_p(t)} \left( n_p(t) + \frac{1}{2} \right) + (2\pi)^d \delta^{(d)}(\vec{p})\, \phi_0^2(t) \, .
\label{eq:correlator}
\end{equation}

\begin{figure}[t]
\vspace*{-4ex}
\hspace*{-5ex}
\includegraphics[scale=0.4]{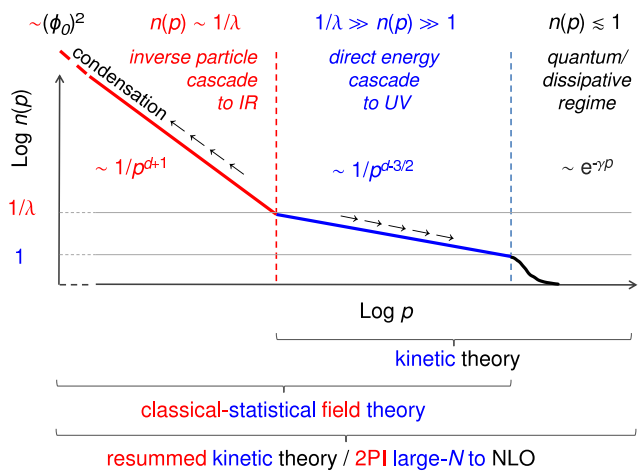}
\vspace*{-8ex}
\caption{Schematic picture of Bose condensation far-from-equilibrium from the dual cascade. Also indicated are the ranges of validity of kinetic theory, classical-statistical field theory and the 'vertex' resummed kinetic theory based on the two-particle-irreducible (2PI) large-$N$ expansion to NLO.}
\label{fig:schematic}
\end{figure}
The situation for Bose condensation far from equilibrium is summarized schematically in Fig.~\ref{fig:schematic}. Displayed are the characteristic regimes of the distribution as a function of momentum on a double logarithmic scale. At large momenta $p  \gg Q_s$, where occupancies $n(p) \lesssim 1$, a quantum or dissipative regime preempts a power-law behavior. At lower but still large momenta $p \gtrsim Q_s$, where parametrically the occupation number is in the range $1/\lambda \gg n(p) \gg 1$, the energy transport occurs via the direct cascade towards higher momenta. In particular, the power-law exponent $d-3/2$ can be understood as a consequence of the presence of a condensate leading to an effective cubic interaction~\cite{Micha:2004bv}. In the low momentum regime $p \lesssim Q_s$ occupancies grow non-perturbatively large, i.e.~parametrically $n(p) \sim 1/\lambda$. Here the inverse particle cascade leads to condensation at $p=0$, which is indicated schematically on this logarithmic plot.    

We emphasize that condensation from initial over-population is not described by conventional kinetic theory including a finite number of elastic or inelastic processes. Instead, an infinite series of processes, all being of order one, has to be summed to get the leading contribution. Fortunately, there exists a non-trivial example where this can be computed directly in quantum field theory. It is based on the two-particle irreducible (2PI) resummed large-$N$ expansion to next-to-leading order for $N$-component scalar field theories~\cite{Berges:2001fi,Berges:2008wm}. It essentially sums elastic and inelastic processes to infinite order as a geometric series, which can be encoded into an 'effective coupling'
$\lambda^{\mathrm{eff}}(p) = \lambda/|1+ \Pi_R(p)|^2$. Here $\Pi_R(p) \equiv \Pi_R(p^0 = p,p)$ denotes the self-consistently dressed (2PI) one-loop retarded self-energy with on-shell frequency $p^0 = p$ \cite{Berges:2010ez}. 
Since $\Pi_R(p \gtrsim Q_s) \lesssim 1$ one finds
\begin{equation} 
\lambda^{\mathrm{eff}}(p \gtrsim Q_s) \simeq \lambda 
\label{eq:pertlambda}
\end{equation}
such that standard kinetic descriptions are recovered at large momenta. In contrast, using the results of Ref.~\cite{Berges:2008wm}, for small momenta $\Pi_R(p) \gg 1$ such that the effective coupling encodes the remarkable infrared scaling 
\begin{equation}
\lambda^{\mathrm{eff}}(p \lesssim Q_s) \, \sim \, \lambda \, 
\left(\frac{p}{Q_s}\right)^8 \, .
\end{equation}
Taking into account also the power-law behavior (\ref{eq:pertscaling}) and (\ref{eq:nonpertscaling}) of the distribution function, scaling analysis shows that elastic processes dominate the regime $p \lesssim Q_s$. The non-condensate fraction of the distribution follows then the effective number conserving 'kinetic equation'
\begin{eqnarray}
\frac{d n_p}{d t} & \sim &  \lambda \int_{l,q,r} {\mathrm{d}} \Omega_{2\leftrightarrow 2} 
\left[ \lambda^{\mathrm{eff}}_{p+l} + \lambda^{\mathrm{eff}}_{p-q} + \lambda^{\mathrm{eff}}_{p-r} \right] \nonumber\\
& \times & \left[ (n_p + n_l) n_q n_r - n_p n_l (n_q + n_r) \right] \, ,
\label{eq:kineticlowp}   
\end{eqnarray}   
where $\int_{l,q,r} {\mathrm{d}} \Omega_{2\leftrightarrow 2}$ denotes the relativistically invariant measure with energy-momentum conservation for two-to-two scattering \cite{Berges:2010ez}. The infrared particle cascade (\ref{eq:nonpertscaling}) is a stationary solution, $d n_p/d t = 0$, of this equation~\cite{Berges:2008wm,Berges:2008sr}.
In the regime $p \gtrsim Q_s$, where (\ref{eq:pertlambda}) holds, the energy cascade  driven by particles scattering off the condensate fraction takes over. The corresponding kinetic description in this momentum regime can be formally obtained from (\ref{eq:kineticlowp}) with the replacement     
\begin{equation}
\frac{n_p}{\omega_p} \, \rightarrow \, \frac{n_p}{\omega_p} + (2\pi)^d \delta^{(d)}(\vec{p})\, \phi_0^2 \, .
\end{equation}     
The energy cascade (\ref{eq:pertscaling}) is a stationary solution of the resulting equation~\cite{Micha:2004bv,Berges:2008wm}.

This scenario of far-from-equilibrium Bose condensation can be rigorously tested using numerical simulations on a space-time lattice. Here we exploit the fact that classical-statistical field theory descriptions~\cite{Epelbaum:2011pc} 
have an overlapping range of validity with
the above effective kinetic theory~\cite{Mueller:2002gd} or the underlying quantum field theory~\cite{Berges:2008wm}. As a consequence, the dynamics of condensation and the relevant cascading regimes with occupancies $n(p) \gg 1$ can be accurately described by numerically solving the classical field equations of motion and Monte Carlo sampling of initial conditions. 

\begin{figure}[t]
\includegraphics[scale=0.7]{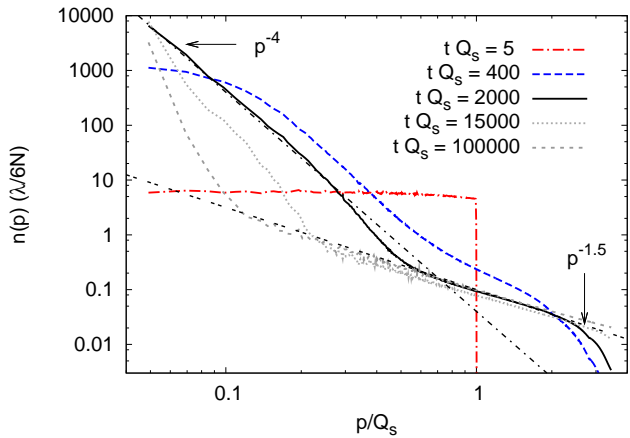}
\caption{Starting from an over-populated initial distribution, a dual cascade develops in the relativistic theory.}
\label{fig:skspektrum}
\end{figure}
\begin{figure}[t]
\includegraphics[scale=0.7]{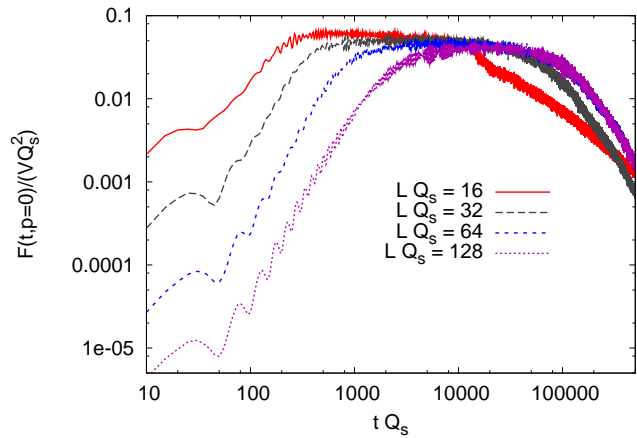}
\caption{Emergence of a volume-independent condensate fraction from runs with different volumes $V=L^3$.}
\label{fig:ccond2}
\end{figure}

In Fig.~\ref{fig:skspektrum} we show the result of simulations for a relativistic $N=4$ component scalar field theory with $\lambda/(4!N) \, \sum_{a=1}^N (\phi_a \phi_a)^2$ interaction in $d=3$. Initial conditions are sampled to generate the non-zero classical averages $\langle \phi \phi \rangle_{\mathrm{cl}} (t=0,p) = n_p^0/\omega_p^0$ and $\langle \dot\phi \dot\phi \rangle_{\mathrm{cl}} (t=0,p) = n_p^0 \omega_p^0$ with initial dispersion $\omega_p = \sqrt{p^2 + m^2}$. Starting from a rescaled distribution $n_p \, \lambda/6N = A \, \Theta ( Q_s - p )$ with amplitudes in the range $A \sim 1 - 10$ to obtain over-population up to the characteristic momentum scale $Q_s$, we find that the system develops the dual cascade with the expected exponents during a long time interval $t Q_s \sim 100 - 10000$. In particular, we find that the effective mass scale $m^2/Q_s^2 \ll 1$ decreases with time while the energy and particle cascades operate. Our main result is given in Fig.~\ref{fig:ccond2}, which shows the zero mode of the correlator $F(t, p=0)/V = \langle \phi \phi \rangle_{\mathrm{cl}}(t,p=0)/V$ as a function of time for different volumes V. For finite $V$ the Dirac $\delta$-function in (\ref{eq:correlator}) at zero spatial momentum is replaced by $(2\pi)^d \delta^{(d)}(0) \rightarrow V$. Fig.~\ref{fig:ccond2} shows that initially this correlator is proportional to $1/V$ since the condensate fraction is zero. The subsequent emergence of a volume-independent contribution signals condensate formation.  
   
Particle number changing processes, which are present in the relativistic theory, finally lead to the decay of the condensate as can be seen from Fig.~\ref{fig:ccond2}. Correspondingly, the energy cascade of the non-condensate fraction shown in Fig.~\ref{fig:skspektrum} moves to the infrared at these times. In theories with exact number conservation a different behavior is expected. For comparison, we consider in the following as a well-known second example the non-relativistic Gross-Pitaevskii equation for scalar fields, which has been studied extensively also in the context of ultracold atomic gases~\cite{BEC}. The relevant coupling $g=4 \pi a / m$ in this case is characterized by the s-wave scattering length $a$ and the mass $m$. Here we employ classical-statistical simulations of that equation to include fluctuations in complete analogy to the relativistic case above~\cite{BS,Scheppach:2009wu}.         
     
\begin{figure}[t]
\includegraphics[scale=0.7]{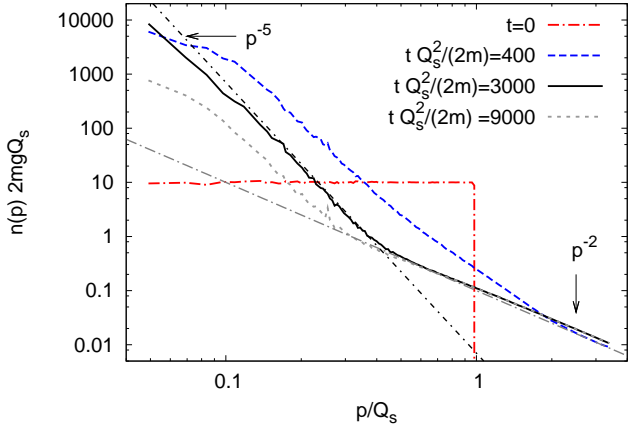}
\caption{Similar to Fig.~\ref{fig:skspektrum}, however, for a non-relativistic theory described by classical-statistical Gross-Pitaevskii theory. The infrared particle cascade emerges with the predicted non-relativistic exponent $d+2$ for the three-dimensional theory.}
\label{fig:gpspect}
\end{figure}

\begin{figure}[t]
\includegraphics[scale=0.7]{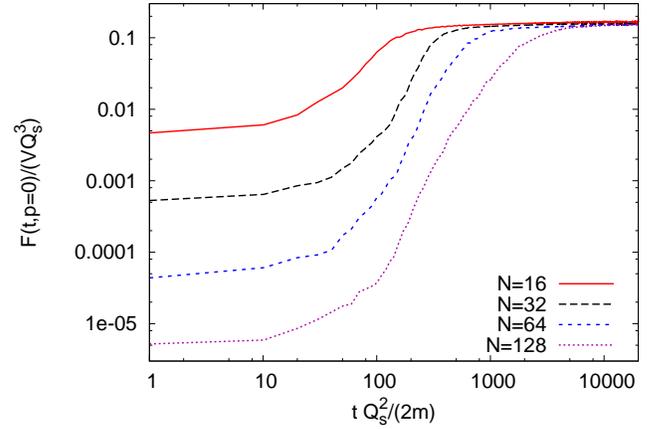}
\caption{Similar as in Fig.~\ref{fig:ccond2}, however, for the non-relativistic theory with conserved particle number. In contrast to the relativistic case, no decay of the condensate due to number changing processes is observed.}
\label{fig:gpcond}
\end{figure}

In Fig.~\ref{fig:gpspect} we show the results for the distribution $n(p)$ in the non-relativistic theory starting from similar initial conditions with over-population as above for $d=3$. Again, the subsequent evolution builds up an inverse particle cascade for $p \lesssim Q_s$, which for the non-relativistic theory is characterized by the scaling exponent $d+2$~\cite{Scheppach:2009wu}. Furthermore, we see that instead of the energy cascade at higher momenta rather quickly a classical thermal distribution $\sim 1/p^2$ occurs. As a consequence of the particle transport towards lower momenta a condensate fraction appears, as is seen from Fig.~\ref{fig:gpcond}. Total particle number $\int \mathrm{d}^d p/(2\pi)^d \, n(p)$ is conserved and we observe no decay of the condensate due to number changing processes.  

We emphasize that the infrared particle cascade is an attractor solution, which is approached from a much wider class of initial conditions than the considered over-population scenario also in different dimensions~\cite{Berges:2008wm,Berges:2010ez}. Its physical nature may be associated to the formation of non-trivial topological configurations, which can be observed from the inhomogeneous field evolution before ensemble averaging~\cite{Scheppach:2009wu,Gasenzer:2011by,BS}. This provides an alternative viewpoint on the same physics as described by the above effective kinetic theory for homogeneous ensemble averages. The latter also allows one to discuss cascades as non-equilibrium renormalization group fixed points~\cite{Berges:2008sr}. 

Though the above scenario of Bose condensation far from equilibrium exhibits universal properties, which apply to a large class of scalar field theories, its generalization to theories with gauge bosons is far from trivial. Taking the relevant example of relativistic, nonabelian $SU(N)$ gauge theories, for the perturbative regime $1/\alpha_s(p) \gg n_s(p) \gg 1$ it has been shown both from classical simulations and resummed perturbation theory that nonabelian gauge theories and scalar theories can exhibit the same class of weak wave turbulence exponents~\cite{Berges:2008mr}. This scaling behavior may be modified at later times \cite{Mueller:2006up}. In the kinetic regime one can study distribution functions which are derived from equal-time correlation functions, for instance, in Coulomb gauge. Using classical-statistical $SU(2)$ gauge theory simulations in $d=3$ starting from over-populated initial conditions, we indeed find the same weak wave turbulence exponent $3/2$ as for the relativistic scalars presented above~\cite{preparation}. 

However, it is unclear of how to interpret the most infrared modes beyond the kinetic regime in this way and suitable gauge-invariant signatures associated to condensation are to be devised. A first non-perturbative study of coherent nonabelian gauge field dynamics is given in Ref.~\cite{Berges:2011sb}. It has also been argued that the infrared cascading solutions observed for scalars~\cite{Berges:2008wm} can be carried over to nonabelian gauge theories~\cite{Carrington:2010sz}. An infrared extended kinetic description as given for the scalars above is complicated for gauge theories, since its straightforward generalization would involve the impractical resummation of all planar diagrams. Classical-statistical simulations, which can also include the relevant physics of longitudinal expansion in the context of heavy-ion collisions, should be the appropriate tool to clarify these questions for sufficiently high $Q_s$.

Acknowledgment: We thank J.-P.\ Blaizot, T.\ Gasenzer, L.\ McLerran, B. Nowak, S.\ Schlichting for very useful discussions. BMBF (06DA9018) and EMMI supported.

\end{document}